\documentclass[12pt]{article}
\usepackage[style=nature]{biblatex}
\usepackage{graphicx}
\usepackage{mathtools}
\usepackage{hyperref}
\usepackage{gensymb}
\usepackage{txfonts}
\usepackage{caption}
\usepackage{authblk}
\usepackage[margin=2.5cm]{geometry}
\usepackage{subcaption}
\usepackage[symbol]{footmisc}
\DeclareUnicodeCharacter{03B3}{$\gamma$}
\begin{document}
\pagenumbering{gobble}
\title{The effect of the electron's spin magnetic moment on quantum radiation in strong electromagnetic fields}

\author{L.A. Ingle$^1$\thanks{louis.ingle@york.ac.uk}, C.D. Arran$^{1}$, M. C. Oxley$^1$, T.G. Blackburn$^2$, S.V. Bulanov$^3$, C.D. Murphy$^1$, C.P. Ridgers$^1$}
\date{}
\affil{\small{$^1$ York Plasma Institute, University of York, Heslington, York, YO10 5DQ}}

\affil{$^2$ Department of Physics, University of Gothenburg, SE-41296 Gothenburg, Sweden}

\affil{$^3$ Extreme Light Infrastructure ERIC, ELI–Beamlines Facility, Za Radnici 835, Dolni Brezany
25241, Czech Republic}

\maketitle
\begin{abstract}
Ultra-intense laser pulses can create sufficiently strong fields to probe quantum electrodynamics effects in a novel regime. By colliding a 60\thinspace GeV electron bunch with a laser pulse focussed to the maximum achievable intensity of $10^{23}$\thinspace Wcm$^{-2}$, we can reach fields much stronger than the critical Schwinger field in the electron rest frame.  When the ratio of these fields $\chi_e\gg1$ we find that the hard ($>25$ \thinspace GeV) radiation from the electron has a substantial contribution from spin-light. 33\% more photons are produced above this energy due to spin-light, the radiation resulting from the acceleration of the electron's intrinsic magnetic moment.  This increase in high-energy photons results in 14\% more positrons produced with energy above $25$\thinspace GeV.  Furthermore, the enhanced photon production due to spin-light results in a 46\% increase in the electron recoil radiation reaction. These observable signatures provide a potential route to observing spin-light in the strongly quantum regime ($\chi_e\gg1$) for the first time.

\end{abstract}
\noindent{\it Keywords\/ QED, Spin, Radiation, Simulation.}
\maketitle
\pagenumbering{arabic}
\section{Introduction}
Recent work has studied the emission of high energy photons and the production of positron-electron pairs in high intensity laser pulses \cite{Bell_08,Ridgers_12,Blackburn_14, Cole_18,Poder_18}. This interaction is often described in terms of synchrotron radiation as the emission from the acceleration of charged particles. Actually the acceleration of a magnetic moment also emits radiation, but most of the time this is much weaker. Now, as experiments reach new record field strengths in intense laser pulses, the magnetic moment due to a particle's spin gains ever more importance. Indeed, the contribution of this 'spin-light' becomes dominant at high field strengths or for strongly relativistic particles. We will explore the effect of 'spin-light' from a particle's intrinsic magnetic moment. and how it affects photon emission and electron-positron pair production.

High-intensity lasers are capable of generating relativistic plasma containing strong electromagnetic fields \cite{Piazza2012,Gonoskov2022,Fedotov2023}. In these environments, a substantial fraction ($> 10\% $) of the laser energy can be converted to gamma ray photons (i.e. $> 10$\thinspace MeV energy) \cite{Nakamura_12,Ridgers_12,Ridgers_13,Brady_13,Ji_14,Hadjisolomou2023}.  This occurs by non-linear inverse Compton scattering (NLCS), i.e. the scattering of many relatively low-energy ($\sim$\thinspace eV) photons by a single energetic electron resulting in a photon of much higher energy \cite{Observing_NLCS} \footnote[1]{Although it should be noted that the distinction from other types of radiation, such as synchrotron emission, becomes increasingly unimportant at high laser intensity \cite{Kirk_09,Ridgers_14}.}. We refer to these high energy photons as ``$\gamma$-ray photons" for the remainder of the article. 
At the limit of intensity achievable by current high-intensity lasers strong quantum effects on radiation can potentially be observed in the laboratory for the first time \cite{Blackburn_19,Yakimenko_19,Baumann_19,Marklund_23}.  The strength of quantum effects on NLCS are parametrised by the quantum parameter for the electrons, $\chi_e$ \cite{Ritus_85}, which is given by,

\begin{equation}\label{Exact_chi}
    \chi_e=\frac{E_{RF}}{E_{sch}}=\frac{\gamma|\textbf{E}_\perp+\textbf{v}\times \textbf{B}|}{E_{sch}}.
\end{equation}

\noindent $E_{RF}$ is the electric field strength experienced by the electron in its rest frame and $E_{sch}$ is the critical (Schwinger) field strength \cite{Schwinger1951}, $1.8\times{10}^{18}$\thinspace Vm$^{-1}$. Quantum effects are important when $\chi_e \geq 1$, and the onset of such effects becomes relevant as we approach $\chi_e \approx 0.1$ \cite{Bell_08,Fedotov_10}. As $\chi_e$ increases, so does the importance of the quantum corrections. For a given field strength achievable in the laboratory $E_{lab}$, it is clear that $E_{RF}$ (and so $\chi_e$) can be boosted if the electrons interacting with this field are highly relativistic, as a relativistic boost gives $E_{RF} = \gamma E_{lab}$ for a Lorentz factor of $\gamma$.

The regime where $\chi_e\sim1$ has been accessed in several ways. By propagating 10 GeV electron beams from a conventional accelerator through a crystal, the fields from the atomic nuclei can add coherently to produce the strong field required, if the lattice is oriented correctly \cite{Uggerhoj}. 
High-intensity lasers can also supply the required strong fields.  Here the laser pulse is ultrashort, fs time-scale and focused to a micron-scale spot size.  In this way, the electromagnetic energy density and so the electric and magnetic field strength from the laser become very high. 

The E-144 experiment at the Stanford Linear Accelerator Center first observed NLCS in the collision of a $47$\thinspace GeV electron beam with a laser pulse in the $\chi_e>0.1$ regime \cite{Observing_NLCS}.  As the laser intensity was relatively low compared to that achievable with modern high-intensity lasers, a very high electron Lorentz factor was required to reach $\chi_e>0.1$.  In this case, however, the interaction is only weakly non-linear, with the electrons only interacting with a few ($<10$) laser photons in the NLCS process. The degree of non-linearity is related to the laser strength parameter (normalised vector potential) $a_0=eE_L/m_ec\omega_L$, with $a_0\gg 1$ required for the interaction to be highly nonlinear.  Here, $E_L$ is the laser electric field and $\omega_L$ the angular frequency.  

In the regime where both the quantum parameter $\chi_e$ and nonlinearity $a_0$ are high and the strong fields exist over a large region relative to the photon formation length (as is the case for laser pulses), the NLCS is much more difficult to calculate.  As there are many laser photons involved in the interaction it is impractical to use the standard approach of considering all possible interactions to form the scattering matrix.  Instead, the interaction with the low energy laser photons is subsumed into the propagator and the `strong-field' quantum electrodynamics (QED) framework is used \cite{Ritus_85}.  In addition, each electron undergoes many NLCS emission events within the laser-pulse and electron motion is treated as classical between these emission events \cite{Ridgers_17,Niel_18}.     

This high $\chi_e$, high $a_0$ `strong-field' QED case important in several extreme astrophysical environments.  For example, electrons and positrons accelerated from a pulsar's surface reach $\chi_e\sim1$ in the magnetic field of the pulsar.  These electrons and positrons emit hard photons, which can decay to electron-positron pairs in the pulsar's B-field.  These pairs are then accelerated, emitting more $\gamma$-ray photons and leading to a cascade of electron-positron pair production that is thought to fill the magnetosphere with an electron-positron plasma \cite{Timokhin_10}. In the magnetosphere of a magnetar the magnetic field can even be greater than the critical field $E_{sch}/c$.  In this case $\chi_e\gg1$ and novel effects such as spin-light become important but this regime has yet to be achieved in the laboratory.  With high-intensity laser facilities such as ELI-NP, ELI-BL and CoReLs demonstrating intensities of the order $10^{23}$ Wcm$^{-2}$ \cite{NP_paper,Corels_laser}, strong-field QED in the regime where $\chi_e\gg1$ will soon be realised in the laboratory \cite{Blackburn_19}.

We will consider two important corrections to NLCS when $\chi_e$ is large. The first is the correction to the radiation reaction, the recoil force experienced by the electron due to its emission.  This has been discussed in detail elsewhere \cite{Ridgers_17,Niel_18} and observed experimentally \cite{Cole_18,Poder_18,Wistisen_18,Los_25} and so will not be discussed in detail here. The second corrects for spin effects, which come in two forms which we will denote as `spin-light' and `spin-flips'. Spin-light refers to the radiation emission due to the acceleration of the spin magnetic moment in an external field, analogous to an accelerating charge \cite{Bordovitsyn}.  Spin-flips refers to the fact that in a strong-field QED treatment of NLCS, the spin orientation of the electron may flip \cite{Ternov_95,Seipt_18,King_20,Del_Sorbo_17_spinflip}. The electron population can then self-polarise as a result of its emission. This polarisation occurs due to an asymmetry in the rate of spin-flips in specific field configurations. One such example is the Sokolov-Ternov effect \cite{Sokolov_Ternov,Ternov_95,Del_Sorbo_17_spinflip}.  Here the rate of spin-flip transitions is not the same in a strong magnetic field. The electrons favour spin-flips to a state anti-parallel to the magnetic field and the population will become polarised. A similar effect can occur in strong laser electromagnetic fields \cite{Del_Sorbo_17_spinflip, DelSorbo_18, Seipt_19,Li_19,Chen_19,Wan_20}.  The rate of NLCS emissions where a spin-flip occurs is generally much smaller than the rate where it does not \cite{Del_Sorbo_17_spinflip}.  The effect of spin on the radiated $\gamma$-ray photon spectrum will therefore be dominated by spin-light.




Spin-light has been observed experimentally in the propagation of a highly relativistic electron beam through crystals as described above \cite{Kirsebom_spin_crystals,Anderson_crystals}. Furthermore, spin-light has been examined by Bordovitsyn et al \cite{Bordovitsyn} and observed in electron storage rings. In both cases the quantum parameter was less than 10 (much less in the former case).  In this paper, we propose a possible experiment for the observation of spin-light in NLCS in the strongly quantum regime $\chi_e\gg1$.  We consider the collision between an energetic electron bunch with energy $>10$\thinspace GeV and a state of the art high intensity laser pulse with intensity $10^{23}$\thinspace Wcm$^{-2}$.  We will show that $\chi_e\gg1$ is necessary for spin-light to dominate the emitted $\gamma$-ray photon spectrum at high energies.

Spin-flips may occur in laser-electron beam \cite{Li_19,Seipt_19,Chen_19,Qian_23}, laser-plasma \cite{GONG_2021} and electron beam-plasma interactions \cite{GONG_2023}. 
In laser pulse-electron beam interactions, spin dynamics can generate highly polarized electron beams which, as shown by simulations, may be achieved using parameters available at current generation laser facilities \cite{Li_19,Seipt_19}. 
Chen \emph{et al.} demonstrated spin dynamics in a laser-electron beam collision results in a highly polarised positron beam source. 
Moreover, an electron beam colliding with two laser pulses may provide a framework for investigating strong-field QED with spin and polarization effects \cite{Qian_23}. 
The resulting spin polarization of electrons that undergo spin-flip transitions in a laser-plasma interaction may act as a spin-based plasma diagnostic to probe the strength and structure of the plasma magnetic field \cite{GONG_2021}. This electron spin polarization has also been used to investigate current filamentation instabilities in beam-electron simulations and may provide an avenue into probing plasma instabilities \cite{GONG_2023}.

While the spin radiation contribution to the radiation spectrum is already included in the radiation models of PIC codes, we aim to examine solely the spin contribution such that we can demonstrate its importance and observability. Thus, our simulations are not adding spin radiation to the existing radiation model, but instead isolating the spin radiation effects and observing it's contribution to the spectrum. 
To demonstrate this, novel particle-in-cell (PIC) simulations using EPOCH \cite{EPOCH2} are performed, the details of which can be found in Section \ref{simulation}. The QED module of EPOCH \cite{Ridgers_14} has been modified such that we can examine a so-called `spinless' case, the details of which are discussed in Section \ref{Theory}.  `Spinless' refers to radiation for which the other relevant corrections have been applied however the spin effects have been subtracted. 
This means we have treated the electrons in this case as scalar particles.  A comparison to this case enables us to isolate the effect of spin radiation\footnote[2]{This is a simpler way to do this than performing the corresponding NLCS calculation in a scalar QED framework and so is our preferred method.}. Our EPOCH simulations use the full quantum, stochastic model of radiation reaction, which is modelled using the locally constant crossed field approximation (LCFA) \cite{Ridgers_17}. The results of our EPOCH simulations will then be presented in Section \ref{results} followed by concluding remarks. 

\section{Spin Radiation}\label{Theory}

The rate of emission of $\gamma$-ray photons by an electron with Lorentz factor $\gamma$ and at $\chi_e$ is given by \cite{Lindhard_91}

\begin{equation}\label{emission_rate}
  \frac{d N_\gamma}{d\chi_\gamma dt} =\frac{4\sqrt{3}\alpha_fc}{3\gamma\lambda_c\chi_e} \left[2K_{2/3}(y)-\int_y^\infty K_{1/3}(t)\thinspace dt\right],
\end{equation}
where $N_\gamma$ is the number of $\gamma$-ray photons, $t$ is the time, $\alpha_f$ is the fine structure constant, $\lambda_c$ is the Compton wavelength and  $\chi_\gamma$ is the photon quantum parameter 

\begin{equation}\label{chi_p_def}
    \chi_\gamma = \frac{h\nu}{m_e c^2}\frac{|\mathbf{E}_{\perp}+\hat{\mathbf{k}}\times c\mathbf{B}|}{E_{sch}}.
\end{equation}
$h\nu$ is the emitted photon's energy and $\mathbf{k}$ its wavevector. $\mathbf{B}$ is the magnetic field experienced by the photon and $\mathbf{E}_{\perp}$ the electric field perpendicular to its trajectory. In the quantum regime, two quantum corrections to equation (\ref{emission_rate}) are required; corrections due to recoil effects and corrections due to spin effects. Equation (\ref{emission_rate}) includes the recoil correction to radiation inside the recoil parameter, $y$. Now using the relation $h(\chi_e)=\frac{1}{\chi_e}F(\chi_e,\chi_\gamma)$, where $F$ is the synchrotron function, we see that
\begin{equation}\label{eq:synch_func_1}
    F_R(\chi_e,\chi_\gamma) = \frac{2\chi_\gamma}{3\chi_e^2}\left[2K_{2/3}(y)-\int_y^\infty K_{1/3}(t)\thinspace dt\right],
\end{equation}
where $F_R(\chi_e,\chi_\gamma)$ is the synchrotron function with recoil effects and the recoil parameter $y$ is given by 
\begin{equation}\label{y}
    y=\frac{2\chi_\gamma}{3\chi_e(\chi_e - \chi_\gamma)}.
\end{equation}
Using the Bessel identity $\left[2K_{2/3}(y)-\int_y^\infty K_{1/3}(t)\thinspace dt\right]=\int_y^\infty K_{5/3}(t) \thinspace dt$, and $(1-\frac{\chi_\gamma}{\chi_e})y=\frac{2\chi_\gamma}{3\chi_e^2}$, we may write $F_R$ as
\begin{equation}\label{F_R}
    F_R(\chi_e,\chi_\gamma)=\left(1-\frac{\chi_\gamma}{\chi_e}\right)y\int_y^\infty K_{5/3}(t)\thinspace dt.
\end{equation}
We have rearranged $F_R$ in this way such that it is consistent with EPOCH notation \cite{Ridgers_17}. 
To thus arrive at the fully quantum synchrotron function, $F_Q(\chi_e,\chi_\gamma)$, we need to introduce corrections for the spin magnetic moment effects. These spin corrections change the coefficient of the $K_{2/3}(y)$ term such that \cite{Erber_66,Ritus_85,Baier_98}
\begin{equation}
    2K_{2/3}(y) \rightarrow \left(1-f+\frac{1}{1-f}\right)K_{2/3}(y),
\end{equation}
where $f$ is the transfer fraction. We define the transfer fraction $f$ as the ratio of the photon and electron energies, given by
\begin{equation}
    f = \frac{\chi_\gamma}{\chi_e} \leq 1,
\end{equation}
based on our definition of $\chi_\gamma$ in Equation (\ref{chi_p_def}). 

In this derivation, no spin direction is assumed. 
The spin-light contribution to radiation examined is not alignment dependent \cite{Del_Sorbo_17_spinflip}, with the radiation having the same power regardless of spin orientation. A full discussion of our treatment of spin polarization can be found in section \ref{Discussion}.

We then multiply$\left(1-f+\frac{1}{1-f}\right)$ by $\frac{1-f}{1-f}$ and then simplify to give
\begin{equation}\label{Bessel_expansion}
    \left(1-f+\frac{1}{1-f}\right)K_{2/3}(y)=\left(\frac{2-2f+f^2}{1-f}\right)K_{2/3}(y)=\left(2+\frac{f^2}{1-f}\right)K_{2/3}(y).
\end{equation}
Substituting this into the synchrotron function in equation (\ref{eq:synch_func_1}) yields
\begin{equation}\label{eq:synch_func_2}
    F_Q(\chi_e,\chi_\gamma) = \frac{2\chi_\gamma}{3\chi_e^2}\left[\left(2+\frac{f^2}{1-f}\right)K_{2/3}(y)-\int_y^\infty K_{1/3}(t)\thinspace dt\right].
\end{equation}
We then separate the $K_{2/3}(y)$ terms and rearrange equation (\ref{eq:synch_func_2}) to obtain
\begin{equation}\label{eq:synch_func_3}
    F_Q(\chi_e,\chi_\gamma) = \frac{2\chi_\gamma}{3\chi_e^2}\left[2K_{2/3}(y)-\int_y^\infty K_{1/3}(t)\thinspace dt\right] + \frac{2\chi_\gamma}{3\chi_e^2} \frac{f^2}{1-f}K_{2/3}(y).
\end{equation}
The first term on the right hand side of equation (\ref{eq:synch_func_3}) is equal to the sychrotron function with recoil corrections given in equation (\ref{F_R}). Thus substituting the definition in equation (\ref{F_R}) into equation (\ref{eq:synch_func_3}) we obtain
\begin{equation}\label{F_Q}
    F_Q(\chi_e,\chi_\gamma)= F_R(\chi_e,\chi_\gamma) + \frac{2\chi_\gamma}{3\chi_e^2} \frac{f^2}{1-f}K_{2/3}(y).
\end{equation}
By separating out the contributions as in equation (\ref{F_Q}), we have separated the synchrotron function into the recoil corrected contribution and the remaining spin corrected contribution. 
Hence, we now define $F_Q=F_R+F_S$ where $F_{S}$ is the spin correction to the synchrotron function. Hence, the additional term present in Equation (\ref{F_Q}) when compared to Equation (\ref{F_R}) gives the spin correction,$F_S$, as
\begin{equation}\label{Fspin}
    F_S(\chi_e,\chi_\gamma) = \frac{\chi_\gamma^2}{\chi_e^2}yK_{2/3}(y)=f^2yK_{2/3}(y),
\end{equation}
where we have again rearranged to be consistent with EPOCH notation \cite{Ridgers_17} using $y=\frac{2\chi_\gamma}{3\chi_e^2(1-f)}$. Hence, the fully quantum corrected synchrotron function is the sum of the recoil corrected synchrotron function and the spin contribution  ($F_Q=F_R+F_S$). It is thus straightforward to isolate the spin contribution to the radiation by comparing the spectrum from $F_Q$ to $F_R$. Any differences result from the inclusion or exclusion of $F_S$. 
Note here that we have assumed that the photon formation length is small and so we may make the locally-constant crossed field approximation, in which the spectrum is given by $F_Q$, in which the field changes on timescales much longer than the photon emission.

\begin{figure}[h!]
    \centering
    \includegraphics[width=0.7\textwidth]{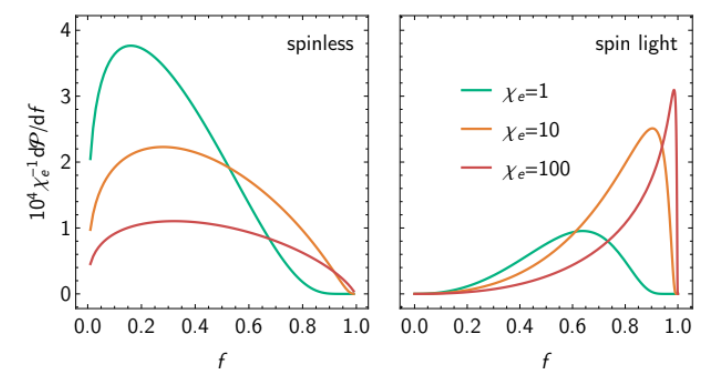}
    \caption{The radiated power with respect to the transfer fraction. Three different quantum parameters of $1$ (Green), $10$ (Orange) and $100$ (Red) are shown. The power has been normalised as shown in order to show all three quantum parameters on the same axis for comparative purposes.  The `spinless' graph refers to the radiation without the spin corrections. The `spin-light' graph is the contribution due to spin. Thus the complete radiation power spectrum is the sum of these two graphs.}
    \label{spin_mag}
\end{figure}

The need to reach $\chi_e >> 1$ to observe spin effects is demonstrated by Figure \ref{spin_mag}, which shows the normalised radiated power with respect to the transfer fraction, $f$ for three different orders of magnitude of $\chi_e$. 
Comparing the spin-light contribution shown on the right in Figure \ref{spin_mag}, to the spinless contribution, shown on the left, we see that for $\chi_e > 10$ spin-light dominates the high energy portion of the spectrum. 

\begin{figure}[h!]
    \centering
    \includegraphics[width=0.7\textwidth]{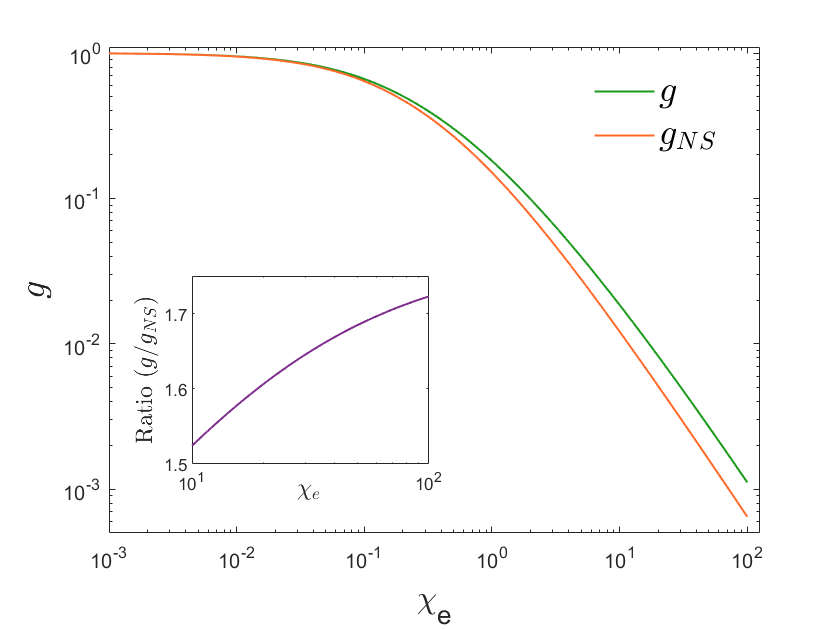}
    \caption{The Gaunt factor with spin (Green) and without spin (Orange). Both the Gaunt factor with and without spin are calculated from their respective fits given by Equation (\ref{gspinfit}) and Equation (\ref{g_ns_fit}) respectively. There is a significant difference due to the inclusion of spin, which grows as a function of $\chi_e$. The figure inset shows the ratio $\frac{g}{g_{NS}}$.}
    \label{gvsgNS}
\end{figure}

We can extract the spin contribution to the radiation reaction from the Gaunt factor, $g$. The Gaunt factor is defined as the ratio of the radiated powers in the quantum and classical cases, and is given by
\begin{equation}
    g  =\frac{3\sqrt{3}}{2\pi \chi_e^2}\int_{0}^{\chi_e /2} F_Q (\chi_e,\chi_\gamma)\thinspace d\chi_\gamma.
\end{equation}
So we can define the spin contribution to the Gaunt factor, $g_{spin}$, as the ratio of the spin radiation power to the total power, such that
\begin{equation}\label{g_spin}
   g_{spin} = \frac{3\sqrt{3}}{2\pi \chi_e^2} \int_{0}^{\chi_e /2} F_{S} (\chi_e,\chi_\gamma) \thinspace d\chi_\gamma.
\end{equation}
Thus, we can express the total Gaunt factor as the sum of these contributions as
\begin{equation}
    g = g_{spin} + g_{NS},
\end{equation}
where $g_{NS}$ is the spinless Gaunt factor;

\begin{equation}\label{g_ns_bessel}
    g_{NS}=\frac{3 \sqrt{3}}{2\pi \chi_e ^2} \int_{0}^{\chi_e /2} F_{R} (\chi_e,\chi_\gamma)\thinspace d\chi_\gamma.
\end{equation}
For simplicity we use the known fit for $g$ \cite{Baier1991_BKS}.
\begin{equation}\label{gspinfit}
    g \approx [1+4.8(1+\chi_e)\ln(1+1.7\chi_e)+2.44\chi_e ^2]^{-2/3},
\end{equation}
and can apply a similar fit to the spinless Gaunt factor.
\begin{equation}\label{g_ns_fit}
    g_{NS} \approx [1+8.754(1+0.6978\chi_e)\ln(1+1.002\chi_e)+5.784\chi_e ^2]^{-2/3}.
\end{equation}
In the $\chi_e\gg1$ limit we can approximate $g_{NS}$ as\footnote[3]{We use the limit of the modified Bessel function $\mbox{lim}_{x\rightarrow0} [K_{2/3}(x)]=\frac{\Gamma(5/3)}{2}\left(\frac{2}{x}\right)^{5/3}$ and the standard integral $\int_0^{\infty}{\frac{x^{1/3}}{(1+x)^2}dx = \frac{2\pi}{3\sqrt{3}}}$, obtained using contour integration by choosing a `keyhole' contour to avoid the branch-cut along the positive real axis in the complex plane.} 
\begin{equation}
    g_{NS} \approx \frac{\Gamma(\frac{5}{3})}{2\sqrt[3]{3}} \; \chi_e ^{-4/3}.
\end{equation}
The comparison between $g$ and $g_{NS}$, using these fits, is shown in Figure \ref{gvsgNS}.

Consider energy conservation for the emitting electron;

\begin{equation}\label{Gamma_cl}
    \frac{d\gamma}{dt}=-\frac{P_{cl}g}{m_e c^2}.
\end{equation}
where $P_{cl}$ is the classical radiated power given by 
\begin{equation}
    P_{cl}=\frac{2 \alpha_f c}{3\lambdabar_c}m_e c^2 \chi_e ^2.
\end{equation}
where $\lambdabar_c$ is the reduced Compton wavelength.
By substituting $g$ in Equation (\ref{Gamma_cl}) for our definition of $g_{spin}$ from Equation (\ref{g_spin}), we arrive at the following expression for spin contribution to the evolution of the electron Lorentz factor,
\begin{equation}\label{gamma_spin}
    \frac{d\gamma_{spin}}{dt}= -\frac{2 \sqrt3 c}{\lambda_c} \int_{0}^{\chi_e /2} f^2y K_{2/3}(y) \thinspace d\chi_\gamma.
\end{equation}
The evolution of the electron Lorentz factor in the spinless case is therefore given by
\begin{equation}\label{dgdt_NS}
\frac{d\gamma_{NS}}{dt} = \frac{d\gamma}{dt} - \frac{d\gamma_{spin}}{dt}.
\end{equation}

\section{Simulations}\label{results}
\subsection{Simulation methodology \& setup}\label{simulation}
The simulations presented here were completed using the of a version of the PIC code EPOCH \cite{EPOCH,EPOCH2} with a modified approach to NLCS.
This approach has been described extensively elsewhere \cite{Nerush_11,EPOCH,Gonoskov_14} but the important aspects of the numerical model will be described here for convenience.  Equation (\ref{emission_rate}) is solved using a Monte-Carlo algorithm to determine when $\gamma$-ray photons are emitted by electrons.  These photons are then added to the simulation and treated as macro-particles whose energy is randomly sampled from $F_Q$, i.e. Equation (\ref{F_Q}).  On modifying EPOCH we introduce the option to also solve the spinless case described above.  The numerical implementation is identical to the spin case just described, but simply replacing $F_Q$ by $F_Q-F_{S}$ in Equations (\ref{emission_rate}) and (\ref{F_Q}). The pair production mechanism in our simulations is non-linear Breit wheeler pair production. In both the spin and spinless cases the NLCS photons can decay to electron-positron pairs in the laser electromagnetic fields via the non-linear Breit-Wheeler process. The model for this in EPOCH is identical in both cases and described in Ref. \cite{EPOCH}. 
This means we do not treat electrons and positions as spinless when it comes to pair creation. 
Pair production is important as it is most probable at higher photon energies and so can reduce the number of high energy photons and so, as we will see, the signature of spin-light.

The 1D EPOCH simulations were set up as follows. Simulations were performed in 1D using a laser with an intensity of $10^{23} \; $Wcm$^{-2}$ colliding with a counter-propagating electron bunch with $60\;$GeV average energy and a Gaussian distribution with 21 \thinspace MeV energy spread (defined as the standard deviation of the distribution). The laser intensity we use is achievable at current generation facilities, for example at ELI-NP, ELI-BL and CoReLs \cite{NP_paper,Corels_laser}. The maximum electron energy currently available from current generations facilities is $\approx 50$\thinspace GeV at SLAC \cite{SLAC_FACET_II_CITATION,E_144_Bamber_99} so our electron energy is slightly larger than the current capabilities. We chose this energy as a next generation parameter to demonstrate the most distinct difference due to spin. Next generation facilities will hopefully have both the desired laser intensity and electron beam energy at the same facility. 
Two different temporal profiles were considered for the laser pulse.  An effectively square profile (i.e. the laser reaches maximum intensity almost instantaneously) was used to simplify the analytical solution to Equations (\ref{Gamma_cl}) \& (\ref{gamma_spin}).  A more realistic Gaussian temporal profile with temporal duration of $10$ fs was also used. We discuss the justification for this laser pulse temporal duration in Section \ref{Discussion}.
In both cases, the laser wavelength was $800 \; $nm. 
The electron density in both cases was a Gaussian profile with $2\; \mu $m FWHM, with a peak density of $3\times10^{11} \; $m$^{-3}$, with a total charge of $7.08\times10^{14}$\thinspace C. This charge is small but since the electron behaviour is single particle this can easily be scaled to pC. 
This was discretised using $2\times10^5$ macro-electrons.  A $50 \; \mu$m length domain was simulated, comprising of $1000$ grid points. As spin-light is most apparent at high photon energy, we set a minimum energy for the $\gamma$-ray photons of $1 \; $GeV to alleviate the computational cost.$\gamma$-ray photons emitted with energy below this energy contribute to radiation reaction but are not simulated as macro-particles. The total time for the simulation was $200 \; $fs.  We also consider cases with pair production artificially switched off as well as cases where it is included.  We therefore consider two cases, increasingly more realistic and complex.  These are denoted as follows: 1) A simpler case using a square laser temporal profile\footnote[7]{The laser temporal profile uses a semi-Gaussian profile with a short rise time to give the closest approximation to a square pulse while suppressing numerical ringing. The semi-Gaussian function is given by
\begin{equation}
  f(t) =
  \begin{cases}
    A\; \text{exp}(-((t-t_0)/\omega)^2) & t<t_0 \\
    A & \text{otherwise}
  \end{cases}
\end{equation}
where $A$ is the maximum amplitude and $t_0$ is the time at which the Gaussian would reach its maximum value.  We find that $t_0=8$\thinspace fs is a good compromise between a rapid rise time and suppressing ringing.}, pair production switched off; 2) A more realistic case utilising a Gaussian laser temporal profile and with pair production switched on. In all cases a circularly polarised laser pulse is used for simplicity (as this removes the laser-cycle dependence of $\chi_e$ present for linearly polarisation). While the results would not be expected to be different with linear polarisation, we do note that linear polarization could reach a higher peak $\chi_e$ \cite{Grismayer2017}. 

\subsection{Simpler case -- square laser temporal profile, pair production switched off}\label{Without Pair Production}
\begin{figure}[h!]
    \centering
    \includegraphics[width=0.7\textwidth]{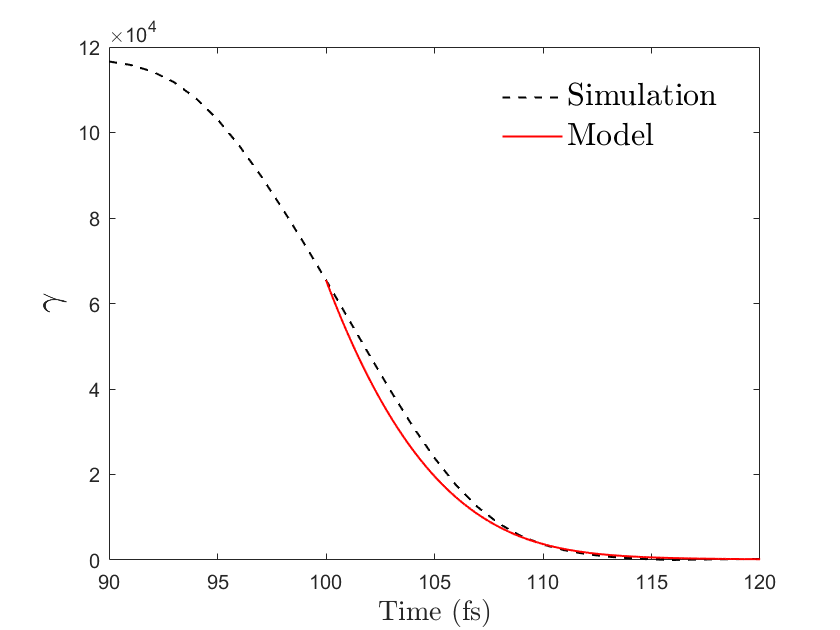}
    \caption{The average Lorentz factor of the electrons with respect to time for the collision of the electron bunch with a counter-propagating laser pulse with a square temporal profile. The simulation data (Purple) is compared to the prediction from Equation (\ref{dgdt_NS}) (Red). The start point for the model of $100$\thinspace fs was chosen to be sufficiently after the rise of the laser, where the laser is close to the peak laser intensity.} 
    \label{semigauss_gamma}
\end{figure}

To investigate the effects of spin on the radiation spectrum, we examine the energy spectrum, beginning with the simple case 1.  The evolution of the average Lorentz factor of the electrons, in the spinless case, is shown in Figure \ref{semigauss_gamma}.  We may straightforwardly solve Equation (\ref{dgdt_NS}) as the laser pulse is circularly polarised and the temporal envelope is a square pulse and thus the electric and magnetic fields are of constant magnitude.  We see that the simulation agrees well with the solution to Equation (\ref{dgdt_NS}) as expected. The agreement is not exact, however this is also expected. $d\gamma/dt$ for the model is for a single particle whereas for the simulation we examine $d\gamma_{ave}/dt$. Thus, the simulation data is averaged over the electron distribution which we would not expect to be identical to the single particle behaviour.

\begin{figure}[h!]
    \centering
    \includegraphics[width=0.7\textwidth]{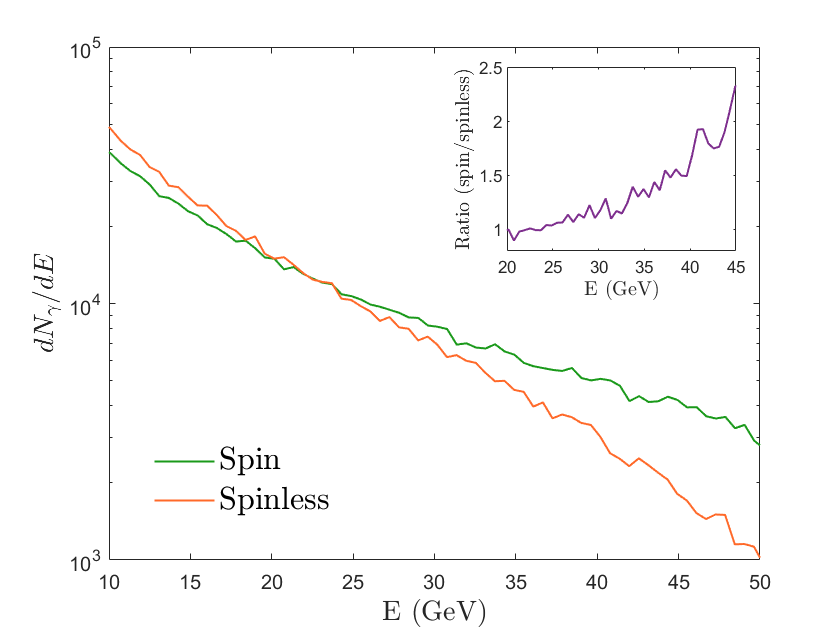}
    \caption{The photon spectrum at $180$ fs when spin radiation is included (Green) and when it is excluded (Orange) for a laser with a square temporal profile. The inset shows the ratio of $(dN_\gamma/dE)_{spin}$ to $(dN_\gamma/dE)_{spinless}$.}
    \label{Semi_gauss_dis_func}
\end{figure}

The photon energy spectrum after 180\thinspace fs is shown in Figure \ref{Semi_gauss_dis_func}. From Figure \ref{semigauss_gamma} we see that by this time the electrons have lost most of their energy and so the instantaneous level of NLCS has dropped to the point where we can consider the interaction to be complete.  Figure \ref{Semi_gauss_dis_func} clearly shows that spin-light clearly dominates the high- energy part of the radiation spectrum ($\gamma-ray$ photons with energy $>25$\thinspace GeV), recalling that spin-flip events are only expected to make a small contribution.

\subsection{More realistic case -- Gaussian temporal laser profile, pair production switched on}\label{With Pair Production}
Next, we shall examine a more realistic case using a Gaussian temporal envelope and including pair production. Using a Gaussian pulse, as would likely be the case in experiment, means that the electrons radiate and lose energy before reaching the peak laser intensity.  This acts to reduce the maximum value of $\chi_e$ the electrons can reach \cite{Blackburn_19} and so is expected to suppress spin-light. As for pair production, spin radiation results in more high-energy $\gamma$-ray photons being produced, and these photons are more likely to pair produce, potentially eroding the high-energy part of the radiation spectrum and so the signature of spin-light.   Furthermore, the pairs themselves will radiate numerous $\gamma$-ray photons. As the produced pairs will be at a lower $\chi_e$ value they will radiate fewer $\gamma$-ray photons due to spin-light, but could still provide a background of energetic photons which obscure the signature of spin-light.

\begin{figure}[h!]
    \centering
    \includegraphics[width=0.7\textwidth]{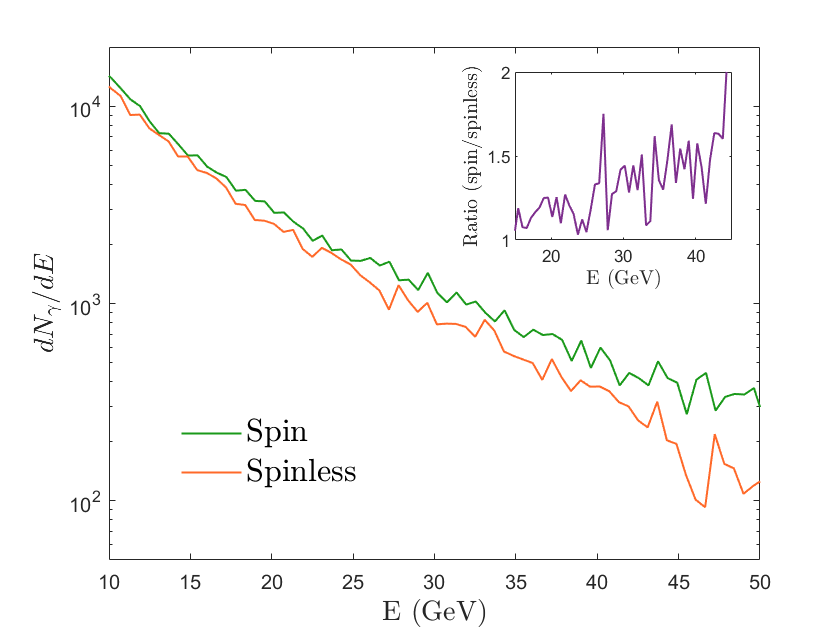}
    \caption{The photon spectrum at $180$ fs with spin (Green) and the spinless case (Orange) for the Gaussian temporal profile. The inset shows the ratio of $(dN_\gamma/dE)_{spin}$ to $(dN_\gamma/dE)_{spinless}$.}
    \label{10fs_pho_dis_180fs}
\end{figure}

Figure \ref{10fs_pho_dis_180fs} shows the spectrum of $\gamma$-ray photons after the interaction of the previously described electron beam with a laser pulse with a Gaussian temporal envelope of FWHM 10\thinspace fs (and pair production included).  We see that spin-light still dominates the high-energy part of the photon spectrum but the signature is not as clear as in the simpler case considered in Section \ref{Without Pair Production}.  We find that there are only $11\%$ more $\gamma$-ray photons produced overall with spin compared to when we neglect spin effects. However, considering the number of $\gamma$-ray photons with energies $\geq 25$ GeV, this difference due to spin radiation effects is greater, with $33\%$ more $\gamma$-ray photons being produced due to spin-light.  

\begin{figure}[h!]
    \centering
    \includegraphics[width=0.7\textwidth]{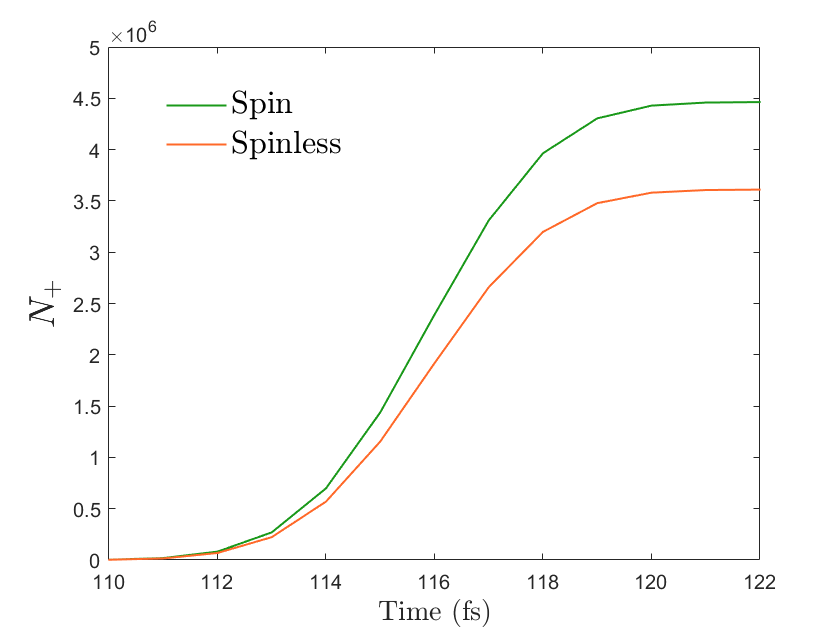}
    \caption{The number of positrons produced with respect to time with spin (Green) and without spin (Orange).}
    \label{positron}
\end{figure}

As electron-positron pairs are produced more favourably by higher energy photons we might expect there to be more produced when spin-light is included, given the results in Figures \ref{Semi_gauss_dis_func} and \ref{10fs_pho_dis_180fs}. The positron yield in the spin and spinless cases is shown in Figure \ref{positron}. When spin-light is included we do indeed see a larger number of pairs produced  when including spin effects.  Note that, due to computational constraints, photons with energy less than 1\thinspace GeV were not included in the simulations but these could generate pairs.  For this reason we looked at the high-energy positrons with $25$ GeV or greater energies.  Here we see a larger difference with $14\%$ more positrons with energies of $25$ GeV or more produced due to the inclusion of spin for the $10$ fs Gaussian laser temporal profile.

\begin{figure}[h!]
    \centering
    \includegraphics[width=0.7\textwidth]{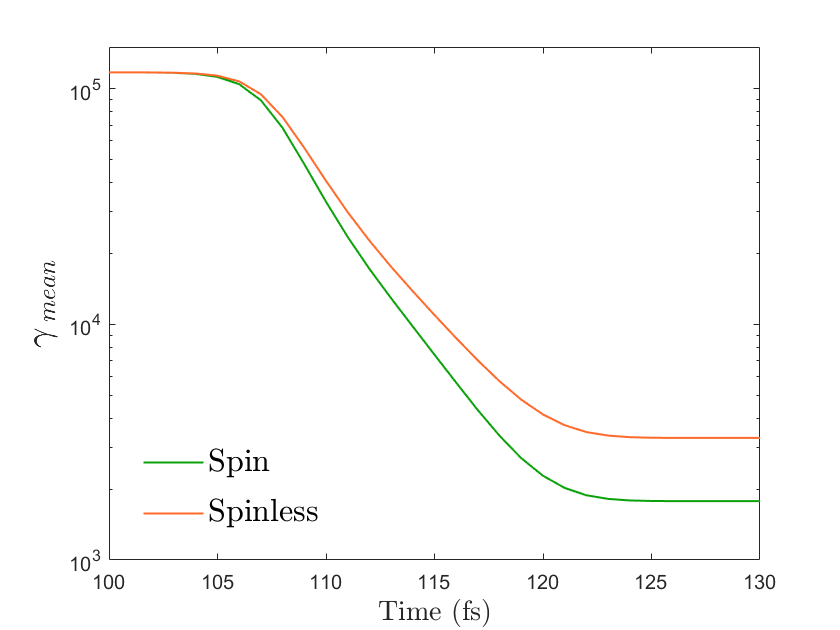}
    \caption{The average Lorentz factor of the electrons with respect to time with spin (Green) and without spin (Orange).}
    \label{gamma_dis}
\end{figure}

Finally, because there is a difference in the energy of the photons radiated we would expect a difference in the corresponding radiation reaction on the electrons. This was examined through the average Lorentz factor of the electrons, as shown in Figure \ref{gamma_dis}.
It is clear that the average Lorentz factor of the electron differs, decreasing more rapidly due to the inclusion of spin radiation. Comparing the electron Lorentz factor after the laser-electron interaction, we see that there is a $46\%$ decrease due to the inclusion of spin radiation effects. When we consider the high-$\chi_e$ limit of $g$ ($(2.44\chi_e^2)^{-2/3}$) and $g_{NS}$ ($(5.784\chi_e^2)^{-2/3}$), the percentage difference between them is $44\%$, thus this difference we observe in the model is reflected in the difference in the Lorentz factor.
\section{Discussion}\label{Discussion}
These results demonstrate that the collision of an energetic electron bunch with a high intensity laser pulse provides a route to studying spin-light in the strongly quantum $\chi_e>>1$, highly non-linear $a_0\gg1$ case. In our more realistic simulation of such an interaction, spin-light results in an $11\%$ difference in the number of $\gamma$-ray photons produced.  A substantially larger (33\%) difference was seen for high energy ($>$25\thinspace GeV) photons. The emission of more high energy photons with the inclusion of spin radiation also alters the radiation reaction on the electrons as demonstrated by Figure \ref{gamma_dis}, with the average Lorentz factor being $46\%$ less after the interaction with the laser pulse when the spin radiation is included. When including non-linear Breit-Wheeler pair production, we observe an increase of $14\%$ in the number of positrons produced, due to increased emission of high energy photons by spin-light. These are signatures of spin-light which could potentially be observed experimentally. 


The pulse duration considered in the more realistic case was $10$ fs which is shorter than is available on existing high-intensity laser facilities. 
We have also conducted equivalent simulations to those presented in Section \ref{With Pair Production}, with more realistic $20$ and $40$\thinspace fs pulse durations.  In these cases the experimental signatures of spin-light are much reduced. 
Alternatively, the pulse can be made effectively shorter duration if the collision is not head-on. For example if the collision between the laser-pulse and the electron beam occurs at 90 degrees, then the duration of the laser field experienced by the bunch is set by the laser focal spot size.  This can be of the order of one micron, which has a light crossing time of 3.3\thinspace fs. This effective reduction in the laser pulse duration can compensate for the loss in peak $\chi_e$ due to the reduced Doppler effect at 90 degrees compared to head-on \cite{Blackburn_19}.

Realising such an experiment necessitates both a highly relativistic electron bunch and high intensity laser. Conventional accelerators such as those at SLAC FACET II \cite{SLAC_FACET_II_CITATION} can provide the appropriate electron bunch while the CoReLs laser \cite{Corels_laser} can provide the required laser intensity. The scarcity of facilities which can provide both has led to all-optical arrangements \cite{Cole_18,Poder_18} despite challenges associated with overlapping the intense laser with a laser wakefield accelerator (LWFA) generated electron beam. 
An example facility which would have access to the required laser intensity and electron beam energies is the European X-ray Free Electron Laser (EuXFEL) \cite{EuXFEL_nature_paper,LUXE_CDR}. The EuXFEL has both the high energy electrons and the ReLaX high intensity laser (although we note this laser does not yet reach $10^{23}$\thinspace Wcm$^{-2}$) making this a potentially ideal set-up for such an experiment. 

\begin{figure}[h]
    \centering
    \includegraphics[width=0.9\linewidth]{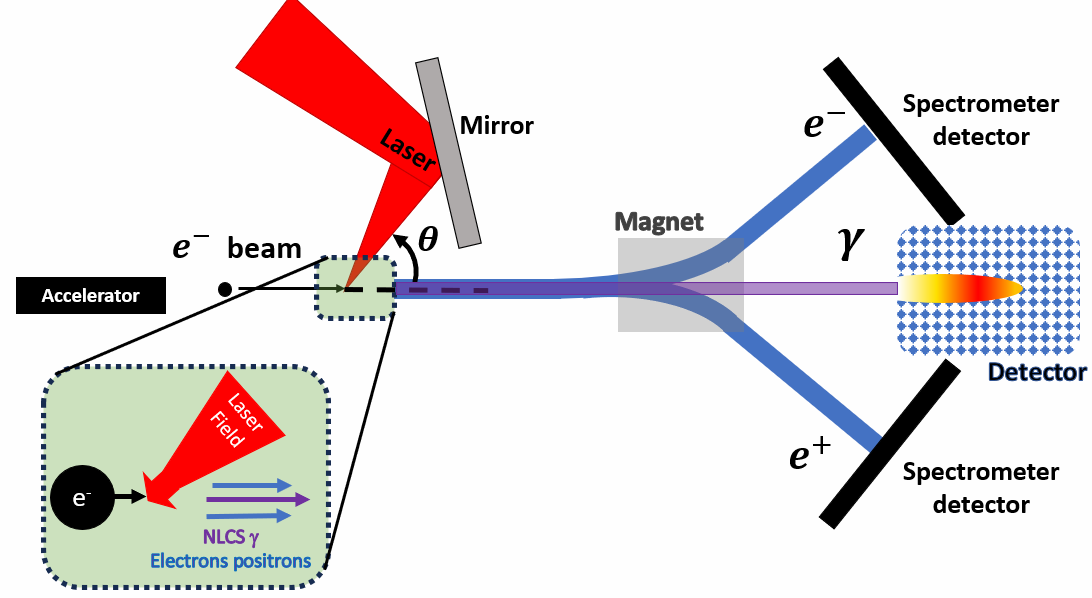}
    \caption{Experimental schematic diagram for a laser pulse-electron beam collision. $\theta$ is the angle between the electron beam and the laser pulse propagation direction. $\theta=0$ represents a head-on collision, whilst $\theta=90\degree$ represents the configuration examined by Blackburn \emph{et al.} \cite{Blackburn_19}.}
    \label{fig:Experiment_Schematic_angled}
\end{figure}

We present a schematic diagram for the potential experimental observation of these spin signatures in Figure \ref{fig:Experiment_Schematic_angled}, based on the design by Cole \emph{et al.} \cite{Cole_18}. $\theta$ is the angle between the electron beam and laser pulse propagation direction, where $\theta = 0$ corresponds to the head-on collision, as in our simulations. The advantage of the counter-propagating configuration is the relativistic boost to the laser fields, and the consequent increase in $\chi_e$. For this collision, equation \ref{Exact_chi} becomes 
\begin{equation}\label{counter_chi}
    \chi_e = \frac{E_{RF}}{E_{sch}}=\frac{2\gamma E_L}{E_{sch}},
\end{equation}
where the factor of $2$ arises due to the relativistic transformation of the laser field from the laboratory frame to the electron rest frame. This gives the largest possible relativistic boost to the laser field and is therefore advantageous for reaching high-$\chi_e$ where the spin effects are most significant.

While the head-on collision provides the greatest relativistic boost, as discussed above, the collision at angle $\theta\neq0$ may be advantageous for reaching high-$\chi_e$. 
The major difficulty with the laser pulse-electron beam collisions at angle $\theta$, as in Figure \ref{fig:Experiment_Schematic_angled}, is the alignment of the laser focal spot with the electron beam. Some temporal and spatial alignment challenges may be mitigated by counter-propagating the beams as in \cite{Cole_18,Poder_18}, which may explain why the $90\degree $ collision is less explored.

The energetic electron bunch may be generated by conventional accelerators \cite{E_144_Bamber_99,SLAC_FACET_II_CITATION} which are capable of creating high quality, more reproducible beams, allowing spin signatures to be more regularly observed. Such beams tend to be much larger than the intense laser focus resulting in only a small fraction of the electrons experiencing the peak of the laser field, with many electrons experiencing a much lower laser field strength. In addition, the timing of the arrival of the electron beam to the laser-pulse may also prove challenging, which may make detecting the spin signature in radiation reaction more difficult. 
However, we would not expect the detection of spin-light to be more difficult as $\gamma$-ray photons will only be produced by the electrons that interact with the laser pulse. 
The main disadvantage to the conventional accelerator is that facilities producing the required electron energy are not in the same location as the required high intensity laser. We noted previously that EuXFEL could be an ideal facility with both a conventional accelerator and high intensity laser in the future. 

An alternative to a conventional accelerator is a LWFA \cite{Tajima_79_LWA_paper}.  
Facilities will often split the laser into several beam lines, one of which can then be used for the LWFA. Hence, there are several laser facilities with the required laser intensity which could provide both the laser pulse and the LWFA driven electron beam at a single location \cite{Corels_laser,NP_paper}. This is beneficial as the LWFA produces an electron beam which is inherently synchronised to the laser-pulse, as both the laser pulse and the LWFA beam originate at the same oscillator.  Laser wakefield acceleration is intrinsically more chaotic due to the plasma dependence meaning there is significant shot-to-shot variance in the electron source \cite{Magnusson_23}. Current generation laser wakefield accelerators may also be unable reach the required electron energies used in our simulations. 

To examine the photon spectrum, we employ a photon detector. Typically, a scintillator crystal is used as the detector \cite{Cole_18,Poder_18,Behm_18}. However, these detectors are used for photon energies of up to $100$ MeV \cite{Behm_18}. Our simulations demonstrate expected photon energies that far exceed this threshold ($E_\gamma \gg 1$ GeV). Thus, an alternative photon detection method would be to use a convertor foil, which has previously detected photon energies up to $150$ GeV \cite{Wistisen_18}. For the electron (positron) detector, we suggest the use of an electron (positron) calorimeter as in the E-144 experiment \cite{E_144_Bamber_99} due to the high energy of the electrons (positrons). 

Our simulation represent the collision of a $60$ GeV electron beam with a $10^{23}$ Wcm$^{-2}$ laser pulse. This corresponds to an approximate peak $\chi_e$ of $77$. Even if this $\chi_e$ was achieved experimentally, be it with an conventional or all-optical set-up, there remains an inherent uncertainty in the experimental measurement of $\chi_e$ for the interaction. For $\chi_e=77$ as in our simulations, the models with and without spin can be differentiated using the electron energy loss and positron production, within a percentage difference of approximately $34\%$ in $\chi_e$, suggesting such an experiment is feasible. 

EPOCH assumes an electron beam with zero net polarization and this beam remains unpolarised throughout the simulation. Even with no net polarization, the spin magnetic moment of an individual electron may radiate via the spin-light. 
Previous work has shown that due to asymmetry in spin transition rates the electron population will polarise over time, however this spin-flip radiation contribution is much smaller than the radiation due to the spin magnetic moment with no spin-flip (i.e. spin-light) \cite{Del_Sorbo_17_spinflip}. Spin light is not alignment dependent. Thus, we anticipate that this interpretation of spin polarisation does not significantly impact the results presented. A full investigation into the effects of spin-flips, such as polarisation and precession, within this framework is beyond the scope of this work but nevertheless provides an opportunity for future research. 

In conclusion, we have used novel 1D EPOCH simulations of a high-intensity laser-pulse colliding with an energetic electron bunch to demonstrate the possibility of experimentally observing spin-light in high $\chi_e$ regime. There is a clear difference in the photon energy spectrum, particularly in the high energy portion, due to spin-light. This results in significant signatures of spin-light in the radiation reaction on the electrons and in pair production.

\section{Acknowledgements}\label{acknowledging}
LAI, CPR and CDM would like to acknowledge that this research was carried out as part of a PhD program funded by the Defence Science and Technology Laboratory on the UK-French PhD Scheme.
CPR, CDA  and CDM would like to acknowledge funding from the UK Engineering and Physical Sciences Research Council, award number EP/V049461/1.  CPR would also like to acknowledge funding from ELI-ERIC.
The Viking cluster was used during this project, which is a high performance compute facility provided by the University of York. We are grateful for computational support from the University of York, IT Services and the Research IT team.
\section{Data Availability Statement}
All the data required to reproduce these results are available at the following URL/DOI: \url{https://doi.org/10.15124/37f506f2-61f2-4087-9714-1492ea384e5e}
\clearpage
\printbibliography
\end{document}